\documentclass[preprint,12pt]{aastex}
\usepackage{epsfig}
\usepackage{rotating}      

\newcommand     \sga  {Sgr~A$^*\:$}

\newcommand     \beq    {\begin{equation}}

\newcommand     \eeq    {\end{equation}}

\newcommand     \gtsim  {\gtrsim}                

\newcommand     \kms    {\,{\rm km~s}^{-1}}
\newcommand     \kpc    {\,{\rm kpc}}

\newcommand     \ltsim  {\lesssim}               

\newcommand     \sm     {M_{\odot}}
\newcommand     \smyr   {M_{\odot}\:{\rm yr^{-1}}}

\newlength{\figwidth}
\countdef\decade=200
\advance\decade by \year
\countdef\hours=201
\advance\hours by \time
\divide\hours by 60
\countdef\mins=202
\advance\mins by \hours
\multiply\mins by 60
\multiply\hours by 100
\countdef\miltime=203
\advance\miltime by \hours
\advance\miltime by \time
\advance\miltime by -\mins

\begin{document}

\title{X-ray Scattering Halos from the Galactic Center:\\
Implications for Diffuse Emission Around \sga\\}

\author{Jonathan C. Tan and B. T. Draine
        }
\affil{Princeton University Observatory, Peyton Hall, Princeton,
NJ 08544; {\tt jt@astro.princeton.edu, draine@astro.princeton.edu}}

\begin{abstract}
  We consider the absorption and scattering of X-rays observed from
  the Galactic center. One objective is to characterize the intrinsic
  X-ray emission from the central black hole, \sga, in its quiescent
  and flaring states --- crucial for our understanding of the
  accretion physics of supermassive black holes. We correct the fluxes
  observed by the Chandra and XMM telescopes for absorption and
  scattering, but limited knowledge about the properties of the
  intervening gas and dust causes large uncertainties. We use
  realistic models for the dust grain size distribution, consistent
  with many other observational constraints, as well as reasonable
  models for the gas and dust abundances and spatial distributions.
  Since much of the intervening dust is relatively close to \sga, the
  scattered halo of X-ray photons is very concentrated: its intensity
  can dominate the point spread function of Chandra inside 1\arcsec,
  and so affects estimates of the point source flux. It also causes an
  apparent broadening of the radial intensity profiles of Galactic
  center sources, and observations of this broadening can therefore
  help constrain models of the line of sight distribution of the dust.
  We estimate that the combined scattering halos from observed
  Galactic center sources within $10\arcsec$\ of \sga contribute up to $\sim
  10\%$ of the observed diffuse emission in this region. Unresolved
  sources may make an additional contribution.  Dust-scattered photons
  suffer a time delay relative to the photons that arrive directly.
  For dust that is 100~pc towards us from the Galactic center, 
  this delay is about
  1000~s at angles of 1\arcsec\, and 100~ks at 10\arcsec. We illustrate
  how the evolution of the scattering halo following X-ray flares from
  \sga or other sources can also help to constrain the dust's line of
  sight distribution. We discuss the implications of X-ray scattering
  halos for the intensity of diffuse emission that has been reported
  within a few arcseconds of \sga: in the most extreme, yet viable,
  model we consider, $\sim 1/3$ of it is due to dust scattering of
  an unresolved source. The remainder results
  from an extended source of emission. 
\end{abstract}


\section{Introduction\label{sec:intro}}
      
Our Galaxy's central supermassive black hole provides an unprecedented
opportunity for studying the accretion physics of these objects, which
are thought to power active galactic nuclei. Since we must view the
Galactic Center (GC) through a large column of gas and dust,
observations are restricted to wavelengths in and longer than the near
infrared (NIR) and to X-rays harder than about 2~keV.

Near infrared observations reveal a dense stellar cluster (Genzel et
al. 2003a, and references therein), whose inner members have high
proper motions, consistent with the presence of a black hole of mass
$\sim 3\times 10^6\sm$ (Sch\"odel et al. 2002; Ghez et al. 2003a). Its
location, as derived from stellar orbits, coincides with the compact,
nonthermal, weakly variable, linearly and circularly polarized radio
source \sga (Balick \& Brown 1974; Bower et al. 2003, and references
therein) and a flaring X-ray point source discovered by the {\it
  Chandra X-ray Observatory} (Baganoff et al. 2001; 2003, hereafter
B01 and B03, respectively). The X-ray emission from this source is
extended on scales comparable to the Bondi radius, $\sim 0.04$~pc.
Recently a flaring NIR source has been detected at this position
(Genzel et al. 2003b; Ghez et al. 2003c).

The spatial and spectral structure of the X-ray emission may help to
constrain the diverse set of theoretical models that have been
proposed for the accretion onto the black hole (e.g., Falcke, \& Markoff 2000; 
Melia, Liu, \&
Coker 2000; Liu \& Melia 2002; Quataert 2002; 
Yuan, Markoff, \& Falcke 2002;
Yuan, Quataert, \&
Narayan 2003; Quataert 2003). It has been argued that the low luminosity of \sga with
respect to its expected classical Bondi accretion luminosity is
evidence for very inefficient accretion mechanisms (e.g. Narayan 2002,
and references therein). Models that attempt to explain the arcsecond
scale diffuse emission via accretion onto a clustered population of
unresolved neutron stars have also been proposed (Pessah \& Melia
2003).  Nayakshin \& Sunyaev (2003) proposed that the X-ray flares may
result from the interaction of stars with a cold gas disk orbiting
close to \sga.

In this paper we consider the scattering and absorption of X-rays from
the Galactic Center. In particular we emphasize the importance of
scattering in producing extended emission --- ``halos'' --- on {\it
  arcsecond} scales. Since much of the intervening dust is located
close to the GC, these halos are much more concentrated than those
around sources observed along more typical Galactic lines of sight. We
assess the contribution made by the concentrated scattering halos of
Galactic center sources to the diffuse emission seen within
$10\arcsec$ of \sga. The halos are so concentrated that they can
dominate the intensity of the Chandra point spread function (PSF)
inside 1\arcsec, so they affect determinations of the point source
flux and thus intrinsic luminosity. They also affect the estimates of
the intensity of emission due to diffuse gas close to the Bondi
radius, and thus estimates of the gas density and the black hole's Bondi
accretion rate.

We consider the distribution of gas and dust towards the Galactic
center in \S\ref{S:gas}. We then use X-ray observations and a model
for how dust and gas absorb and scatter X-rays to calculate the
intrinsic X-ray luminosity and spectrum of \sga (\S\ref{S:spec}).
This involves a correction for dust scattering that requires a
calculation of the angular intensity profiles of scattering halos as a
function of energy.  In \S\ref{S:halo} we calculate the total
intensity of the scattering halo, including examples of the delayed
halos from flares, and compare to observations. We discuss the
implications of our results and conclude in \S\ref{S:con}.

\section{The Gas and Dust Distribution to the Galactic Center\label{S:gas}}

To calculate X-ray scattering we require a model for the spatial
distribution of dust along the midplane of the Galaxy from the Sun to
the Galactic center. The total column of gas and dust also affects the
amount of absorption. We take the distance from the Sun to the
Galactic center to be 8~kpc (Reid 1993). We assume a uniform
metallicity and dust-to-gas ratio, taken to be 1.5 times the value for
the local ISM, consistent with Galactic abundance gradients measured
in \ion{H}{2} regions (Giveon, Morisset, \& Sternberg 2002).

\subsection{Spatial Distribution}

Based on studies of the radial distribution of the atomic and
molecular gas (e.g. Binney \& Merrifield 1998, Fig.\ 9.19) we
approximate the distribution of gas from the main Galactic disk via a
Gaussian located at $R=4$~kpc and with a HWHM of 2~kpc. We denote the
fraction of the total column that is in this outer component by $f_o$.
This approximation underestimates the local ISM density close to the
Sun, but this material makes little contribution to inner X-ray
scattering halos, on scales of arcseconds.

A second component of the total column is the material in the inner
few hundred parsecs of the Galaxy, and we denote the fraction of
material in this region by $f_i$. The properties of this gas have been
constrained by data from IRAS ({\it Infrared Astronomical Satellite})
and COBE ({\it Cosmic Background Explorer}) observations (Launhardt,
Zylka, \& Mezger 2002). A gas disk with typical densities in its
diffuse component of $n_{\rm H} \sim 300\:{\rm cm^{-3}}$ is inferred
to extend out to $R\sim 200-300\:{\rm pc}$. An inner, denser,
inclined, circumnuclear disk is seen as a separate dynamical component
(G\"usten et al. 1987; Marr, Wright, \& Backer 1993) extending from
$R\simeq 2 - 10\:{\rm pc}$. However, most of this material does not
intersect our line of sight to \sga. The distribution of gas in these
regions is very inhomogeneous: within the $\sim 200$ parsec disk are
giant molecular clouds, such as the ``50~km/s cloud'' and the
``20~km/s cloud'', which lie close to the line of sight to \sga.
Vollmer, Zylka, \& Duschl (2003) estimate that the former is $\sim
0-5\:{\rm pc}$ closer to us than \sga, while the latter is spread out
over a larger distance with its densest region $\sim 25-50\:{\rm pc}$
closer than \sga. The centroids of these clouds are within $\sim
5-10\:{\rm pc}$ of our \sga sight line. There is also evidence for
clumps of hot molecular gas in the circumnuclear disk on scales $\sim
10\:{\rm pc}$ (McGary, Coil, \& Ho 2001), and even inside the central
parsec (Herrnstein \& Ho 2002), so that the total gas column to a
particular location in the central region may show significant
variation. In the study of Herrnstein \& Ho (2002) structure was seen
on scales all the way down to the beam size of 10\arcsec$\simeq
0.4\:{\rm pc}$.

As a first attempt to describe this complicated density structure, we
shall approximate the radial distribution of the inner component with
a Gaussian centered at various distances, $R_i$, towards us from \sga,
and with a HWHM equal to $R_i$. As a fiducial case we take
$R_i=100$~pc, but we also consider models with $R_i=50$ and 25~pc.  In
all cases we assume a small central cavity that is free of dust. We
take the radius of this cavity to be 2~pc, and then adjust the
normalizations of the Gaussians so that the total column from 2~pc to
8~kpc is equal to the adopted overall totals.  Individual gas clumps
and clouds may have a column density comparable with the total through
the more diffuse regions (below). To allow for the possibility of such
an enhancement we consider models with a range of column densities in
the smoothly distributed component.\footnote{Note that the scattering
  halos due to material inside the inner few parsecs are typically
  smaller than the size of the point spread function (PSF) of Chandra,
  and thus do not significantly affect the observable halo profile.}

\subsection{Estimates of the Total Column Density}

The total columns of gas and dust to the GC are uncertain. One of the
most common methods used to measure these quantities is by estimation
of the K band extinction by simultaneously fitting near-IR photometry
of GC stars with spectra of a particular stellar type and with a
certain amount of reddening (Becklin et al. 1978; Lebofsky, Rieke, \&
Tokunaga 1982; Sellgren et al. 1987; Rieke, Rieke, \& Paul 1989,
hereafter RRP).  This method requires knowledge of the extinction law
in the near-IR (e.g. Rieke \& Lebofsky 1985) and accurate models for
the intrinsic stellar spectra.  The main discriminant for the stellar
spectra are the strengths of the CO absorption bands: they become
stronger for decreasing temperature and increasing luminosity. $\rm
H_2O$ absorption at the edges of the K~band can help distinguish
between M giants and K supergiants, because the strength of this
absorption decreases with increasing luminosity. Typically, solar
metallicity stars have been used as standards for late-type spectra
(e.g. Kleinmann \& Hall 1986).  One possible source of systematic
error is the effect of elevated metallicities, which would tend to
increase the strength of the CO and $\rm H_2O$ features for a given
luminosity and surface temperature.  

An alternative to fitting detailed stellar spectra is to assume that
certain, particularly blue, sources have Rayleigh-Jeans spectra: this
method was used by RRP for the sources making up the IRS16 complex,
which have subsequently been identified as hot, emission line stars
(e.g. Eckart et al. 1995; Krabbe et al. 1995; these observations also
show that IRS13, discussed below, is a complex dominated by hot
He~$\rm I$ stars), however the derived dust columns are still somewhat
sensitive to the choice of extinction law (see below).

A third method to measure extinction is from the ratio of Br$\alpha$
to Br$\gamma$ recombination lines. The difficulties here are (1) the
need to assume a particular theoretical ratio for the intrinsic line
ratio, e.g. based on case B recombination theory at a particular
density and temperature, and (2) the measurement of the line ratios in
what is a very complicated and crowded region of the sky.

The relatively bright infra-red sources closest to \sga are the IRS16
complex, IRS13, and IRS7, which have projected separations from \sga
of $\sim 0.1$, 0.13, and 0.20~pc.  The values of $A_K$ adopted by RRP
for these sources were 3.4, 3.5, 4.1, respectively, where the value
for IRS16 is an average of the 7 components reported by RRP. The
extinction law used by RRP has $A_\lambda \propto \lambda^{-\beta}$
between $1\mu m$ and $3\mu m$, with $\beta=1.61$.  The Milky Way dust
models of Weingartner \& Draine (2001, hereafter WD01) with $R_V\equiv
A_V/E(B-V)=3.1,4.0,5.5$ have $\beta \simeq 1.73,1.25,1.25$.  If the
extinction at K is derived from the degree of reddening of a
Rayleigh-Jeans spectrum, then the values of $A_K$ derived from WD01
models are $0.93,1.29,1.29$ times that derived by RRP. Similar
corrections may be expected for determinations of $A_K$ from fitting
more realistic stellar spectra.  Draine (2003a) has discussed the
extinction per H nucleon in regions of different $R_V$. Using
$A_I/N_{\rm H}$ from Draine (2003a) and the grain size distributions
of WD01, we find $A_K/N_{\rm H} = 6.04, 7.18, 8.19 \times
10^{-23}\:{\rm mag\: cm^{2}}$.
In the fiducial,
``local Milky Way metallicity'' case of the WD01 model with $R_V=3.1$
the effective H column densities for the three GC sources described
above are thus $5.2, 5.4, 6.3\times 10^{22}\:{\rm cm^{-2}}$.  For the
models with $R_V=4.0, 5.1$ the column densities are different by factors
of 1.09 and 0.92, respectively. If the effective metallicity is
greater than solar by a certain factor, then these estimates would be
reduced by the same factor, assuming a constant dust to metals ratio
and for fixed grain properties.

Stars such as S0-2, which are very close to \sga in both the plane of
the sky and in line of sight dimension, are much fainter than the
brighter IR sources discussed above, so the continuum slopes of their
published spectra (Ghez et al. 2003b) are not yet reliably determined
(A. Ghez 2003, private communication). With this caveat in mind, the K
band extinction can be estimated assuming an intrinsic Rayleigh-Jeans
spectrum (which is a reasonable assumption) and assuming a K band
continuum spectrum $F_\lambda \propto \lambda^\gamma$ with $\gamma=5.0
- 6.5$. These values yield $A_K\simeq 4.6 - 5.4$ for the RRP
normalization, about 50\% greater than the value estimated using the
same technique towards IRS16.

The above range of values for the extinction appear to be fairly
typical of the GC region on larger scales. Using the same reddening
law as RRP, Figer et al. (1998) find $A_K=3.2\pm0.5$ towards the
Quintuplet star cluster, located 30~pc in projection from \sga.
Lebofsky \& Rieke (1987) found the extinction over a 24 square
arcminute region around the GC was spatially variable, but never less
than $A_K \simeq 3.4$. Sellgren et al. (1987) (using $\beta=1.9$ so
that values of $A_K$ are about a factor of 0.85 of those derived using
$\beta=1.61$ of RRP) reported that their extinction results were
consistent with a uniform component with $A_K\sim 2$ and a variable
component due to material near the GC with $A_K\sim 0-1$. These
results are consistent with the recent model of Launhardt et al.
(2002), who find $A_K\simeq 1.7$ ($A_V\simeq 15$) due to dust in the
outer Galactic disk and a similar value for extinction due to dust in
the GC region itself.

We note that the clumps reported by Herrnstein \& Ho (2002) are
estimated (but with large uncertainties) to have $N_{\rm H}\sim
2\times 10^{24}\:{\rm cm^{-2}}$, which would correspond to extinctions
a factor of $\sim$30 larger than typically reported, if they contain
typical dust grain populations. 

As we shall describe later, estimates of the total column density have
also been made by fitting models for the X-ray emission from \sga to
observations. Porquet et al. (2003) found $N_{\rm H}\simeq
1.9 \times 10^{23}\:{\rm cm^{-2}}$, corresponding to $A_K\simeq
11-9.9$ for $R_V=3.3-5.5$. B01 found $N_{\rm
  H}=5.3_{-1.1}^{+0.9}\times 10^{22}\:{\rm cm^{-2}}$, i.e. about a
factor of 3 or 4 smaller than Porquet et al. (2003).

Given the uncertainties in the estimates of the total gas and dust
columns along the line of sight to the GC, and since it is clear that
there is some variation towards different objects in this region, we
shall consider a range of different density distributions and total
column densities.  As an average for the gas along the line of sight
to the GC, we adopt a metallicity of 1.5 times solar and a dust-to-gas
mass ratio of 1.5 times the local interstellar value, which is taken
to be 0.008. This metallicity enhancement is consistent with ISO ({\it
  Infrared Space Observatory}) observations of Galactic \ion{H}{2}
regions (Giveon, Morisset, \& Sternberg 2002).
We assume depletions into dust grains of 67\% for C,
20\% for O, 90\% for Mg, Si, and Fe, and 0\% for N, Ne, and S (WD01).  For
the fiducial model we set $A_K=3.5$, which for the WD01 grain model
with $R_V=3.1$ and the above metallicities yields $N_{\rm H,tot}=3.90
\times 10^{22}\:{\rm cm^{-2}}$. For this model we also set
$f_i=f_o=0.5$. We shall also consider a model with $A_K=7.0$, so that
$N_{\rm H,tot}$ is twice the above value, and which has $f_i=0.75$ and
$f_o=0.25$: i.e. we have added material only to the inner component.

The $R_V=3.1$ dust model with $\bar{Z}=1.5Z_0$
has C/H~=~90~ppm in polycyclic aromatic hydrocarbons (PAHs) and
consists of a mixture of carbonaceous grains and amorphous silicate
grains.  As discussed by Li \& Draine (2001), the carbonaceous grains
have the properties of PAH molecules when they contain $\ltsim 10^4$ C
atoms, and the properties of graphite particles when they contain
$\gtsim 10^5$ C atoms.  By altering the size distributions of the
carbonaceous and silicate particles, the dust model appears to be able
to reproduce observed extinction curves in various Galactic regions,
and in the Large and Small Magellanic Clouds.  This dust model, when
illuminated by starlight, produces infrared emission consistent with
the observed emission spectrum of the interstellar medium (Li \&
Draine 2001, 2002). 

\begin{figure}[h]
\begin{center}
\epsfig{
        file=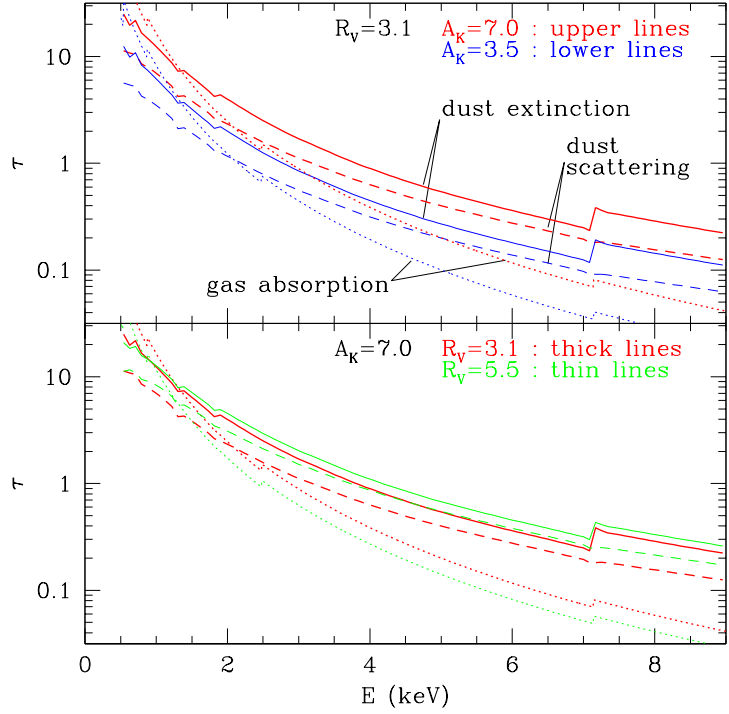,
        angle=0,
        width=4.0in}
\end{center}
\caption{
\label{fig:tau}
\footnotesize Energy dependence of optical depths due to photoelectric
absorption by interstellar gas (dotted lines) and total extinction by
dust (solid lines), including the contribution due to scattering
(dashed lines). The top panel shows models with $R_V=3.1$ and
$\bar{Z}=1.5Z_0$: the lower sets of lines are for $A_K=3.5$ and the
upper for $A_K=7.0$. The hydrogen columns are $N_{\rm
  H}=3.90,7.79\times 10^{22}\:{\rm cm^{-2}}$ respectively.  The lower
panel compares $A_K=7.0$ models that have $R_V=3.1$ and 5.5 (thin
lines). The latter model has $N_{\rm H}=5.48\times 10^{22}\:{\rm cm^{-2}}$.
}
\end{figure}

For the amorphous silicate grains and the carbonaceous grains we
employed the dielectric functions recently estimated by Draine (2003b)
and calculated the scattering using Mie theory.  The X-ray absorption
and scattering properties of this model are consistent with
observations towards Nova Cygni 1992 that probe typical dust in the
local diffuse ISM (Draine \& Tan 2003).  In Figure \ref{fig:tau} we
show the energy dependence of the optical depth (due to photoelectric
absorption by gas and extinction by dust) of our models of gas and
dust toward the GC. The contribution of the dust extinction due to
scattering is also shown.

Lutz et al. (1996) reported extinction towards the GC in the
4-8~$\rm \mu m$ region of the spectrum exceeding that expected for
standard graphite-silicate mixes (e.g. Draine 1989), which may
indicate that additional species are present, such as iron/metal
sulphides or metal oxides, or carriers related to the production of
the 3~$\rm \mu m$ `ice' absorption features that are seen in the spectra.
From an analysis of such absorption features, Chiar et al. (2000)
found evidence for enhanced molecular material along the line of sight
to \sga as compared to the Quintuplet star cluster.

A shift in the grain size distribution toward larger grains, with
increased values of $R_V$, is expected in the dense,
cold conditions of the cores of giant molecular clouds. Since these
conditions may apply to a significant amount of the gas and dust
towards \sga, we shall also consider the effect of using models with
$R_V=5.5$ for the entire dust column.  Dust models with larger grains
yield lower values of $N_{\rm H,tot}$ for a given $A_K$ and
metallicity. For example, the $R_V=5.5$ model of WD01 gives columns
that are a factor of 0.70 smaller than the $R_V=3.1$ case.

Given an observed X-ray spectrum we can use the models shown in Figure
\ref{fig:tau} to estimate the form of the intrinsic spectrum
(\S\ref{S:spec}).  However, we shall also see that when the flux from
a particular source is defined as all photons arriving from within a
certain angular radius (e.g. on arcsecond scales for Chandra), then a
substantial fraction of the photons that are scattered can remain in
this source region and must be accounted for.

\section{The Intrinsic X-Ray Spectrum of \sga\label{S:spec}}

Utilizing the high spatial resolution ($\sim 0.5\arcsec$) of {\it
  Chandra}, B01 and B03 reported the first detection of an X-ray
source that can be convincingly associated with \sga. B03 fit the
observed spectrum (derived from counts within 1.5$\arcsec$ of the
source from the first 40.3~ks observation) with an absorbed power-law
model, with $(d\dot{N}/dE)_0 = A_0 (E/{\rm keV})^{-\Gamma}\:{\rm
  ph\:cm^{-2}\:s^{-1}\:keV^{-1}}$ with $\Gamma=2.7^{+1.3}_{-0.9}$ and
$N_{\rm H} = 9.8^{+4.4}_{-3.0} \times 10^{22}\:{\rm cm^{-2}}$ (90\%
confidence interval). However, these calculations did not account for
dust scattering (B01), leading to an overestimation of the column
density necessary to produce the observed soft X-ray cutoff.  B01
reported results from a second observation with effective exposure
time of 35.4~ks, which included a $\sim 10$~ks flare. From a joint
analysis of the quiescent emission from the two epochs and the flare
emission from the second epoch, B01 derived
$\Gamma_q=2.2^{+0.5}_{-0.7}$, $\Gamma_f=1.3^{+0.5}_{-0.6}$, and
$N_{\rm H} = 5.3^{+0.9}_{-1.1} \times 10^{22}\:{\rm cm^{-2}}$, where
the subscripts $q$ and $f$ refer to the quiescent and flaring states,
respectively. Models of the spectra of an absorbed optically thin
thermal plasma (Raymond \& Smith 1977) yielded similar column
densities with $kT = 1.9^{+0.9}_{-0.5}\:{\rm keV}$ (B03). Systematic
errors due to the energy dependence of the enclosed energy fraction
inside 1.5$\arcsec$ and of charge transfer inefficiency effects could
lead to an overestimation of $\Gamma$ by $\sim 0.2-0.3$, an
underestimation of the intrinsic X-ray luminosity by $\sim 20\%$, and
an overestimation of $N_{\rm H}$ (B03).

XMM-Newton observed a bright 3~ks flare from \sga (Porquet et al.
2003) and from these observations were derived $\Gamma_f=2.5\pm0.3$,
and $N_{\rm H} = 20\pm 3 \times 10^{22}\:{\rm cm^{-2}}$. The spectrum
could also be fit with models based on bremsstrahlung continuum,
thermal blackbody, and optically thin plasma emission.

The method of estimating the intrinsic spectrum of a source typically
involves counting all the photons arriving inside a given circular
area around a source (a radius of $1.5\arcsec$ was used for the
Chandra observations of the quiescent state), and then subtracting off
a background flux, estimated from a surrounding annulus.
Given an assumed
dust and gas column, a correction is then made to estimate the
intrinsic flux that would be received if there was no absorption.
However, for gas and dust distributed close to the source, much of the
scattering halo is located very close to the PSF, and needs to be
accounted for.

\begin{figure}[h]
\begin{center}
\epsfig{
        file=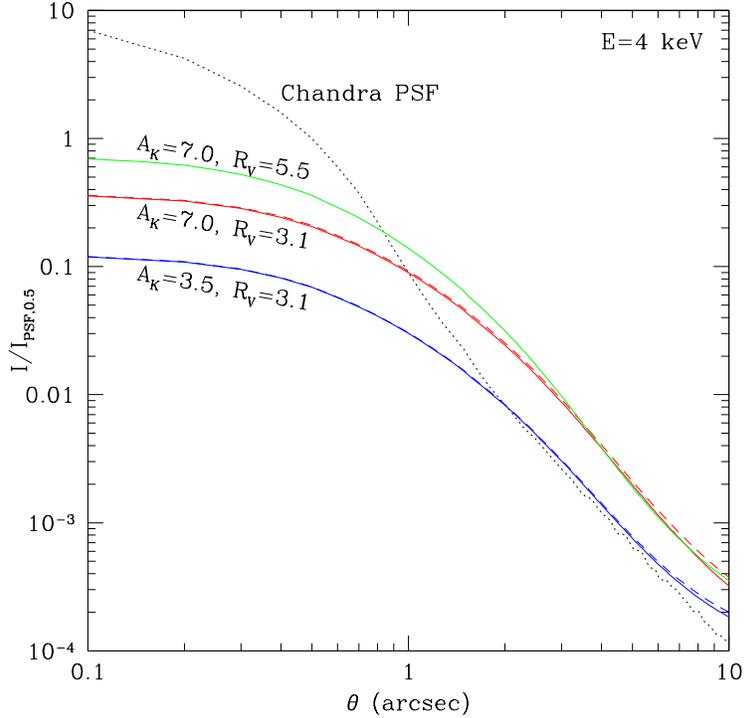,
        angle=0,
        width=4in}
\end{center}
\vspace*{-2.em}
\caption{
\label{fig:halo4}
\footnotesize Radial intensity profile of dust-scattered halos at
4~keV for the models with $A_K=3.5,7.0$
compared to the Chandra PSF (dotted). Note the halos have been
convolved with the Chandra PSF. The solid lines are halos calculated
with single scattering, and the short-dashed including double
scattering. Also shown is the single scattering halo for
the $A_K=7.0$ model with dust that has $R_V=5.5$.}
\end{figure}

\begin{figure}[h]
\begin{center}
\epsfig{
        file=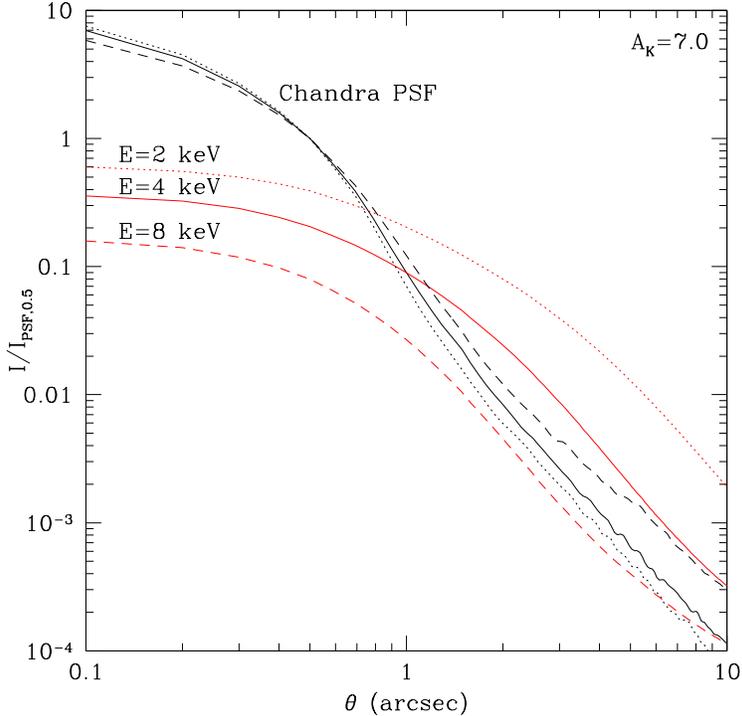,
        angle=0,
        width=4in}
\end{center}
\vspace*{-2.em}
\caption{
\label{fig:halo248}
\footnotesize Radial intensity profile of dust-scattered halos
(convolved with the Chandra PSF) at $E=2,4,8$~keV (dotted, solid,
dashed lines, respectively) for the models with $A_K=7.0$ and
$R_V=3.1$, compared to the Chandra PSF at these energies.  }
\end{figure}

\begin{figure}[h]
\begin{center}
\epsfig{
        file=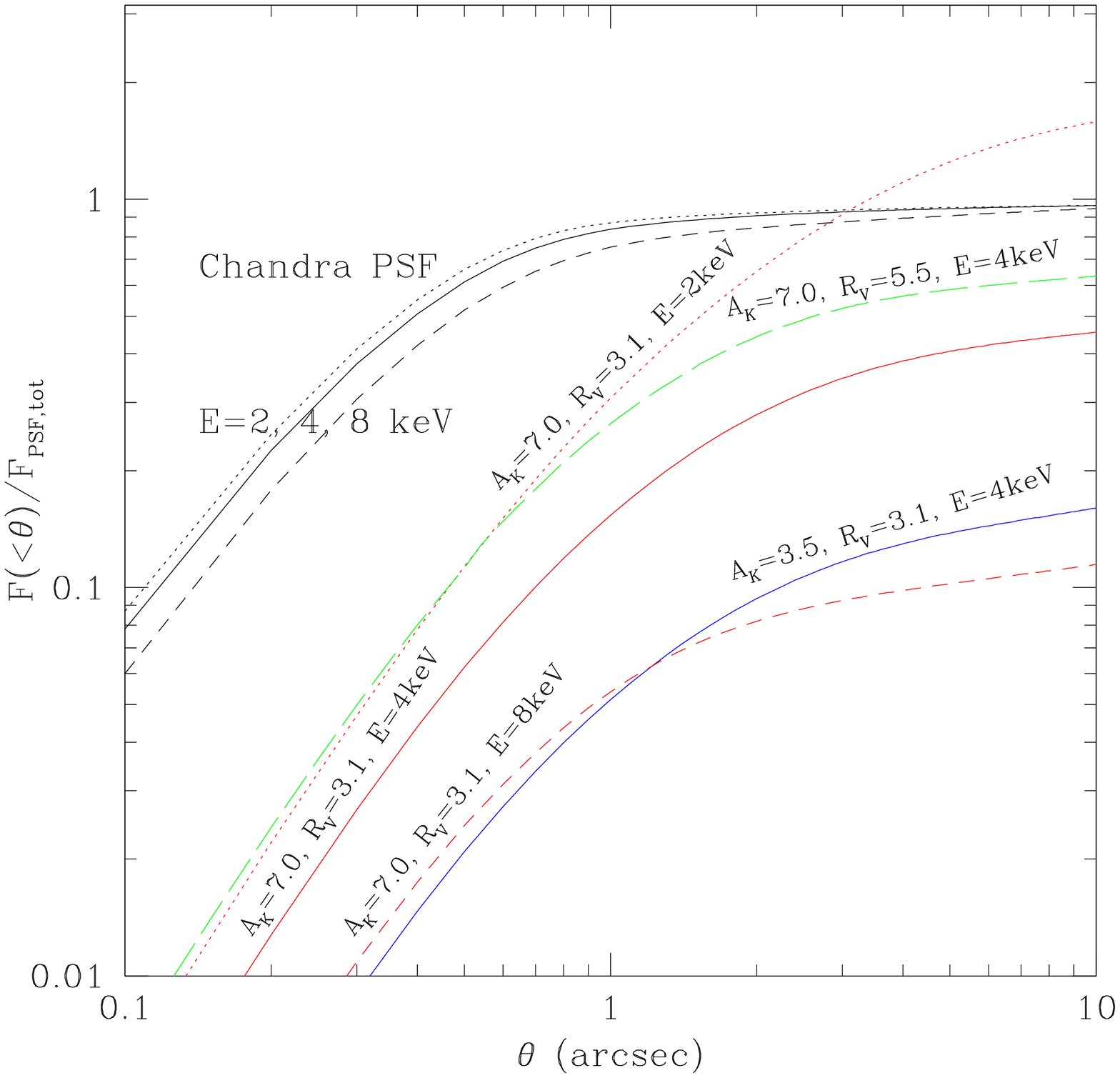,
        angle=0,
        width=4in}
\end{center}
\vspace*{-2.em}
\caption{
\label{fig:haloint}
\footnotesize Radial profile of enclosed fluxes at 2,4,8~keV (dotted,
solid, dashed), for the PSF and single scattering dust halos, with
properties as labeled.  Note the normalization is to the total PSF
flux, so that these functions are equivalent to $(F_{\rm
  halo,tot}/F_{\rm PSF,tot}) g(\theta)$, where $g(\theta)$ is the
enclose halo fraction (see Draine 2003b).
The numerical data defining these functions, with finer energy
resolution and over a broader energy range, are available at
www.astro.princeton.edu/~jt/.
}
\end{figure}

Consider a source observed at 4~keV. For the model with $A_K=3.5$ that
has $f_o=f_i=0.5$, we have $\tau_{\rm gas}=0.19$, $\tau_{\rm
  dust,ext}=0.45$, and $\tau_{\rm dust,sca}=0.32$. Figure
\ref{fig:halo4} shows the radial intensity profiles of dust-scattered
halos of a monochromatic source at 4~keV for the adopted models for
the density distribution to the GC, i.e. for $A_K=3.5$ and $7.0$. The
halo profile was convolved with the on-axis energy-dependent Chandra
PSF, derived from the CIAO (Chandra Interactive Analysis of
Observations) software. We see that the scattering halos can dominate
over the PSF inside 1\arcsec. In this figure we also show the
dependence on the scattering halo on the grain size distribution, as
parameterized via $R_V$: a model with a larger value of $R_V$,
corresponding to larger average grain sizes, produces a halo that is
more concentrated inside a few arcseconds.  The halos at several
different energies are compared in Figure~\ref{fig:halo248}: at lower
energies the halo intensities are stronger relative to the PSF.

The fraction of enclosed flux in the scattering halo, relative to that
in the PSF is shown in Figure~\ref{fig:haloint}. These results allow
one to correct for the contribution that a scattering halo makes to
the direct flux from a source, given the angular scale inside which
the direct flux of a particular observation is estimated, i.e. the
observed flux inside an angle $\theta$ is the sum of the unscattered
flux and the part of the scattered flux that lies inside $\theta$.

\begin{figure}[h]
\begin{center}
\epsfig{
        file=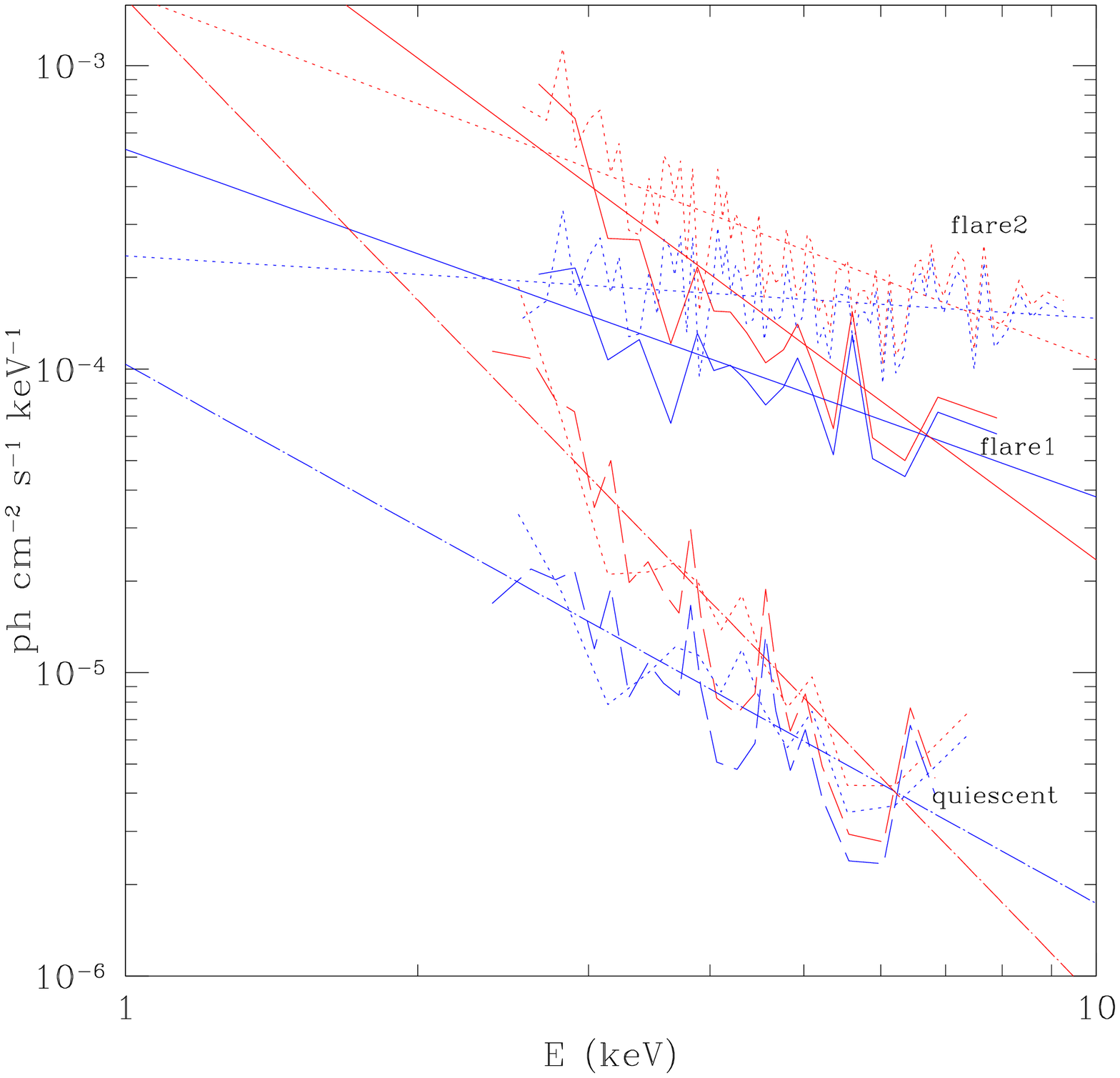,
        angle=0,
        width=4.3in}
\end{center}
\vspace*{-2.em}
\caption{
\label{fig:specqpownew}
\footnotesize Intrinsic, unabsorbed spectra of quiescent and flaring
emission from \sga.  The data (connected, jagged lines) for the
quiescent emission were reported by B01 and B03 (lower sets of dotted
and dashed lines, respectively).  ``Flare1'' was reported by B01
(solid lines) and ``flare2'' by Porquet et al. (2003) (upper dotted
lines). Each data set has been corrected for absorption and scattering
for the $A_K=3.5$ (blue) and $A_K=7.0$ (red) models. In each case the
straight line shows the best power law model (parameters listed in Table \ref{tab:spec}).
}
\end{figure}

For absorption and scattering with our fiducial density models, we
calculate the intrinsic spectra of \sga in its quiescent and flaring
states, based on the observed fluxes and appropriate extraction radii
(Figure~\ref{fig:specqpownew} and Table~\ref{tab:spec}).  From these
results we see that the uncertainty in the intrinsic X-ray luminosity
of \sga is set by the uncertainty in the extinction, as parameterized
by $A_K$. This also sets the uncertainty in the slope of the spectrum,
e.g. when fit by power-law models. The quiescent state 2-10~keV
luminosity is a few$\times10^{33}\:{\rm ergs\:s^{-1}}$. The two flares
reported by B01 and Porquet et al. (2003) are about a factor of 10 to
30 times more luminous in this energy range, when averaged over the
flare durations.  We estimate that the peak luminosity of the flares
is a factor of 1.3 and 2 times greater than the mean values quoted in
the table. We tend to favor a harder intrinsic spectrum for the flare
than Porquet et al., who find their best fitting power-law model has
$\Gamma=2.5\pm0.3$ (and $N_{\rm H}=2\times 10^{23}\:{\rm cm^{-2}}$),
however at least part of this discrepancy is due to the fact that our
model uses a much smaller column of gas: $3.9,7.8\times 10^{22}\:{\rm
  cm^{-2}}$ for the $A_K=3.5,7.0$ cases respectively. Part of this
difference is due to our use of abundances that are a factor of 1.5
times greater than the local interstellar values, and part may be due
to differences in the grain models, including our preference to
normalize to the observed extinction in the K band, rather than the
canonical visual extinction of 30~mag. Our results are broadly
consistent with those of B01.

\begin{deluxetable}{ccccccc} 
\tablecaption{Parameters of Power Law fits to Intrinsic X-ray Spectra of \sga\label{tab:spec}}
\tablewidth{0pt}
\tablehead{
\colhead{State of} & \colhead{$A_K$} & \colhead{$R_V$} & \colhead{$\theta_{\rm ex,sc}$\tablenotemark{b}} & 
\colhead{$A_0$ ($10^{-4}{\rm ph\:cm^{-2}s^{-1}keV^{-1}}$)} & \colhead{$\Gamma$} & \colhead{$L_X(2-10\:{\rm keV})$}\\
\colhead{\sga [Ref.]\tablenotemark{a}} & \colhead{(mag)} & \colhead{} & \colhead{(arcsec)} & 
\colhead{at 1 keV} & \colhead{} & \colhead{($10^{33}\:{\rm ergs\:s^{-1}}$)}
}
\startdata
Quiescent [1] & 3.5 & 3.1 & 1.5 & 1.04 & 1.78 & 2.88\\
Quiescent [1] & 7.0 & 3.1 & 1.5 & 16.6 & 3.30 & 5.58\\
Quiescent [1] & 7.0 & 5.5 & 1.5 & 13.0 & 3.15 & 5.27\\
Flaring [1] & 3.5 & 3.1 & 2.5 & 4.82 & 1.10 & 40.1\\
Flaring [1] & 7.0 & 3.1 & 2.5 & 43.2 & 2.25 & 58.8\\
Flaring [1] & 7.0 & 5.5 & 2.5 & 34.4 & 2.13 & 56.4\\
Flaring [2] & 3.5 & 3.1 & 2.5\tablenotemark{c} & 2.37 & 0.21 & 95.1\\
Flaring [2] & 7.0 & 3.1 & 2.5\tablenotemark{c} & 17.4 & 1.21 & 120\\
Flaring [2] & 7.0 & 5.5 & 2.5\tablenotemark{c} & 14.1 & 1.10 & 117\\
\enddata
\tablenotetext{a}{\footnotesize References: 
(1) Baganoff et al. (2001);
(2) Porquet et al. (2003)}
\tablenotetext{b}{\footnotesize Effective extraction radius used to define source spectrum.}
\tablenotetext{c}{\footnotesize The extraction radius was $10\arcsec$, but from time delay considerations and the duration of the flare $\sim 3000$~s we expect the scattering correction to be limited to that appearing inside $\sim 2.5\arcsec$.}
\end{deluxetable}



\section{The X-ray Scattering Halo of \sga\label{S:halo}}

X-ray scattering by dust creates diffuse ``halos'' of emission around
astronomical sources (Overbeck 1965; Martin 1970; Hayakawa 1973), and
these can be used to constrain models of the dust (e.g. Catura 1983; Mathis, \& Lee 1991;
Predehl \& Klose 1996; Smith, Edgar, \& Schafer 2002; Draine \& Tan
2003, and references therein).  X-ray scattering is most efficient at
small angles, $\lesssim 1$~degree, and if the dust and gas have a
fairly uniform distribution along the line of sight to a source, then
this angle is also representative of the size of the observed
scattering halo. The diffuse emission directly associated with \sga
extends only over a few arcseconds (B03). It has been argued that any
diffuse emission created by dust-scattering would be fairly uniform on
these scales, and therefore removed during the process of background
subtraction: B03 estimated the background from the mean intensity
between scales 3.5\arcsec to 10\arcsec.

However, dust that is close to the source tends to produce a more
intense and concentrated scattering halo. This is relevant to the case
of \sga because much of the line-of-sight dust is thought to be within
a 100~pc or so of the Galactic center. The median scattering angle,
$\theta_{s,50}\simeq 360\arcsec({\rm keV}/E)$, and dust at a distance
$R$ from a point source produces a scattered halo with 50\% of the
scattered power within $\theta_{h,50}\simeq (R/D)\theta_{s,50}\simeq
2.25\arcsec (R/50{\rm pc})(8{\rm kpc}/D)({\rm keV}/E)$, where $D$ is
the distance from the observer to the point source, and multiple
scattering is neglected (Draine 2003).

\begin{figure}[h]
\begin{center}
\epsfig{
        file=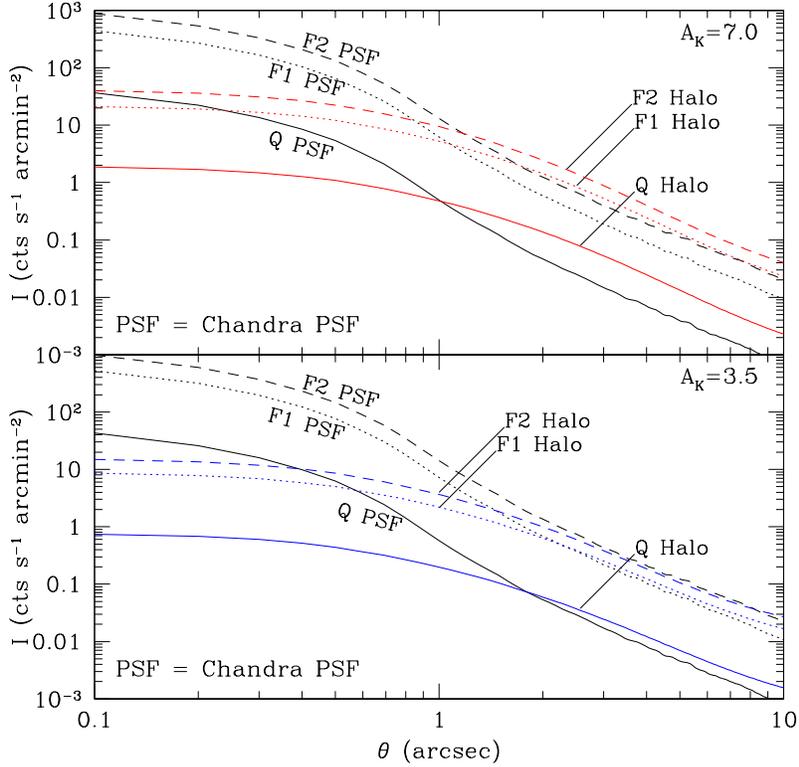,
        angle=0,
        width=4.3in}
\end{center}
\vspace*{-2.em}
\caption{
\label{fig:halologxpsfqf}
\footnotesize Upper panel: Radial intensity profile of dust-scattered
halos (convolved with the Chandra PSF) for the quiescent (solid), and
flaring (dotted, ``F1'' - B01; short-dashed, ``F2'' - Porquet et al.  2003) states of
\sga assuming $A_K=7.0$ and evaluated between 0.5 and 7~keV. The PSF
profiles are also shown.  Note that these ``flaring'' models make use
of the spectra appropriate for the flaring state, but assume a steady
flux (see \S\ref{S:tdelay} for time-varying models).  Lower panel: As
above, but for $A_K=3.5$.
}
\end{figure}

For the adopted models of the source spectrum in the quiescent and
flaring states, Figure~\ref{fig:halologxpsfqf} shows the calculated azimuthally-averaged intensity
profile of the scattering halo. As in
the monochromatic case, the halo profile was convolved with the
on-axis Chandra PSF, derived from the CIAO data analysis
software.\footnote{Smith et al. (2002) have shown that the PSF at
  angles greater than $\sim 10\arcsec$ is best derived empirically from 
  study of astronomical sources. In this study we restrict our
  attention to scales inside $10\arcsec$, where the PSF is well-determined
  from theoretical modeling of the telescope's optics.} Note that the energy
dependence of the PSF can be seen in Figure~\ref{fig:halo248}.

The strength of the scattering halo is relatively more important for
the quiescent state, because of its softer spectrum. However, in all
cases the halo intensity can become significant relative to the PSF on
scales of about an arcsecond or so.

\begin{figure}[h]
\begin{center}
\epsfig{
        file=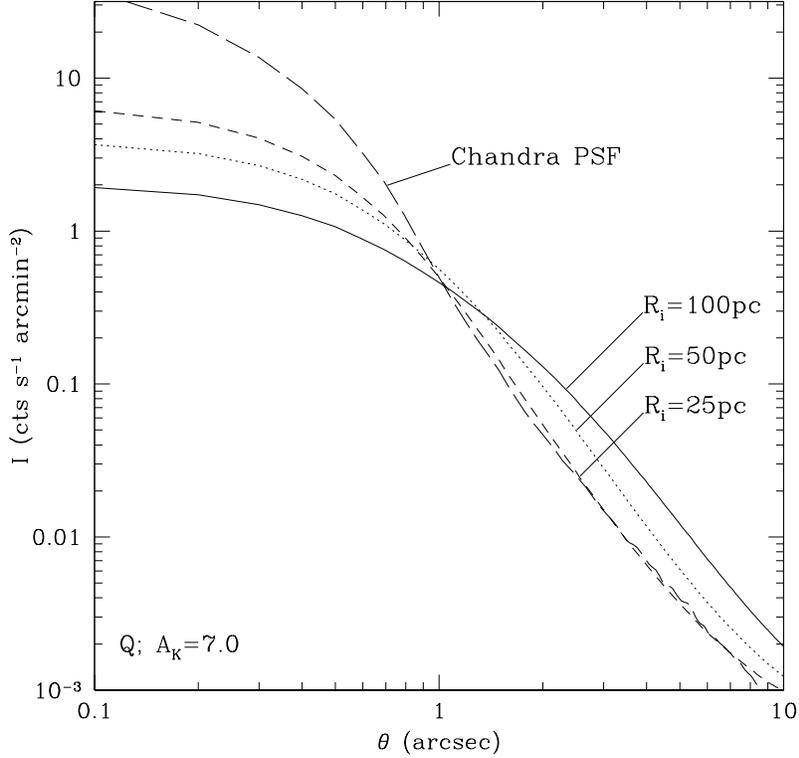,
        angle=0,
        width=4.3in}
\end{center}
\vspace*{-2.em}
\caption{
\label{fig:halologxpsfqfden}
\footnotesize Dependence of radial intensity profile of dust-scattered
halos on the location of inner dust cloud relative to \sga,
illustrated with the quiescent model with $A_K=7.0$ and evaluated between 0.5 and 7~keV.
}
\end{figure}

In Figure \ref{fig:halologxpsfqfden} we show the effect of changing
the location of the inner component of dust that is near \sga. As the
dust is brought closer to \sga, the halo becomes more concentrated,
becoming more and more point-like, and thus indistinguishable from the
intrinsic source after convolution with the (Chandra) PSF.

\begin{figure}[h]
\begin{center}
\epsfig{
        file=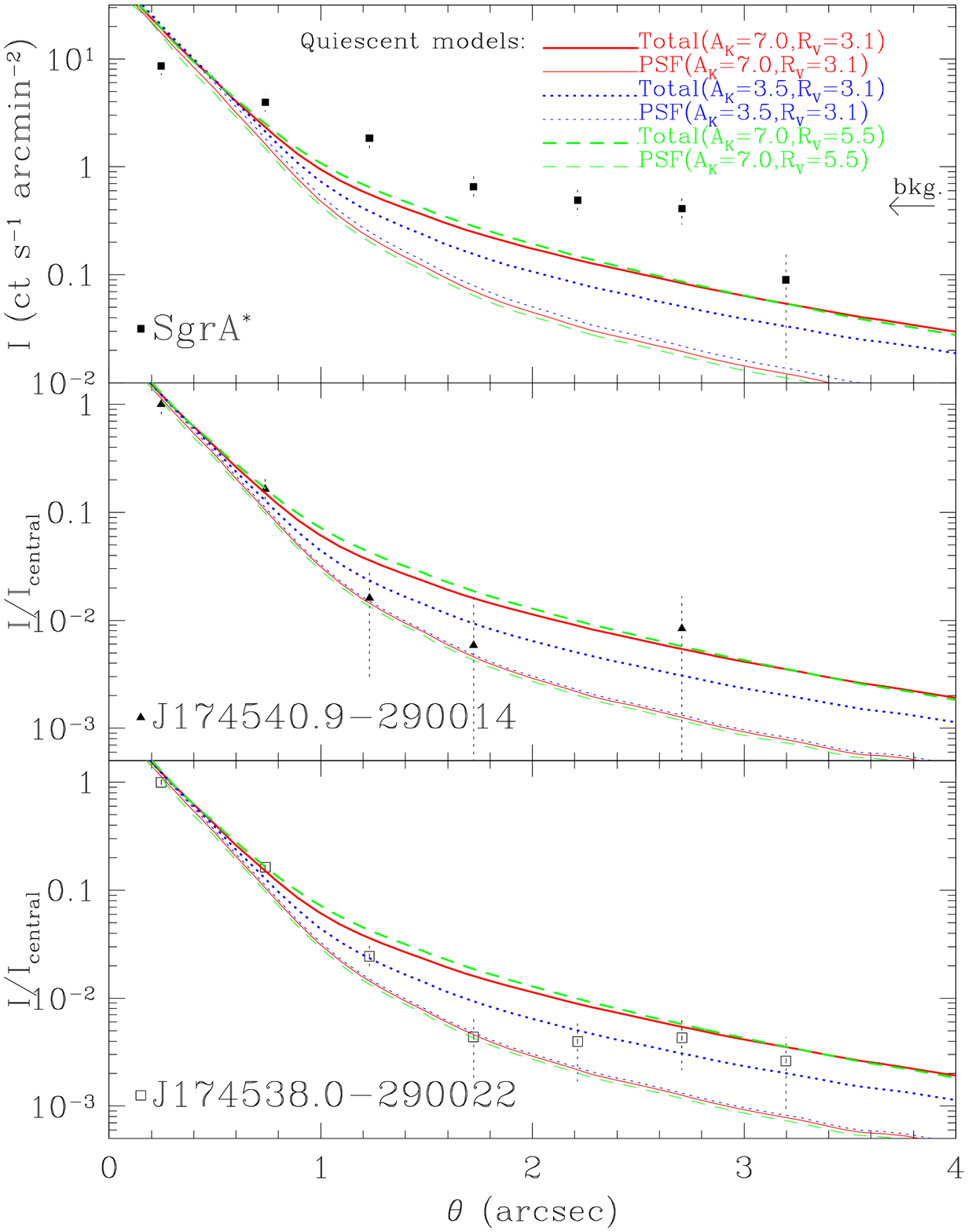,
        angle=0,
        width=4.3in}
\end{center}
\vspace*{-2.em}
\caption{
\label{fig:halonormsimp}
\footnotesize Upper panel: Comparison of observed (B03) radial
intensity profiles of \sga and several theoretical profiles of the
emission (0.5 to 7~keV), assuming it arises from an unresolved, steady
point source, that is convolved with the Chandra PSF, and broadened by
the effects of dust-scattering.  Data are averaged in
Chandra-pixel-wide annuli ($0.492\arcsec$).  The arrow on the right of
the figure shows the intensity of the background that B03 estimated
from 3.5-10\arcsec and subtracted to yield the \sga data shown here.
Two other GC sources are shown in the lower panels (B03). We have normalized
models and data relative to the intensity of the central annulus. Note
that J174538.0-290022 is brighter and softer than \sga, while
J174540.9-290014 is of similar brightness and harder.
The model intensity profiles are
somewhat more concentrated than the observed profile of \sga, but are
in reasonable agreement with the profiles of the other two sources.
This suggests that the general broadening of GC sources noted by B03
is largely due to dust-scattering halos. For this effect to explain
the observed profile of \sga using quiescent models, larger
extinctions are required, approximately 2 or 3 times that of the
$A_K=7.0$ model.  }
\end{figure}

In Figure \ref{fig:halonormsimp} we compare the theoretical radial
intensity profiles, which are the sum of the PSF and the
dust-scattered halo, with the observed emission from and around \sga
in its quiescent state (B03). 
The observations are 
presented as the mean values of annuli 0.492\arcsec wide, which is the
pixel size. Note that the dithering of the telescope during the
observation means that these profiles are not broadened significantly
by the finite pixel size. B03 also corrected for pointing errors,
estimated to be $\simeq0.15\arcsec$ by $\simeq0.37\arcsec$.
We also show the radial intensity profiles
of two other nearby point sources, J174540.9-290014 and
J174538.0-290022, observed by B03, normalized to the intensity of
the innermost annulus.  These sources have comparable spectra to \sga,
are separated on the sky from \sga by 18.7\arcsec (0.73~pc) and
27.0\arcsec (1.05~pc), respectively. Being heavily absorbed they are
likely to be in the GC region, and given the angular distribution of
sources around \sga, their proximity makes it likely that they are
physically close to \sga in all three spatial dimensions. It is
likely, although not certain (\S\ref{S:gas}), that their lines of
sight are subject to similar extinctions, particularly for
contributions from gas and dust that is $R\gtrsim 25\:{\rm pc}$
towards us from the Galactic center.

We find that dust-scattering of point source emission is important for
broadening the PSF of Chandra for typical sources in the vicinity of
the GC. The level of enhancement is sufficient to explain the observed
profiles of the sources near \sga.  The largest enhancement at the
arcsecond scale occurs with the inner dust component about 50~pc from
\sga, and with dust that has $R_V=5.5$, i.e. larger grains.  However,
even with the model with the largest dust column, corresponding to
$A_K=7.0$~magnitudes, the scattering halo appears to be insufficient
to explain the intensity of the diffuse emission within a few
arcseconds of \sga by factors of several.  The total dust column
would have to be increased by similar factors above the value of the
$A_K=7.0$ model in order to explain the entire diffuse emission as
being due to dust-scattered emission from an unresolved, steady point source.

We note that our model intensity profiles are greater than the data in
the central annulus by about a factor of 1.8. This is because we have
normalized the models to produce the observed flux, which was
estimated inside an extraction radius of $1.5\arcsec$. The models
underpredict the intensity beyond about 0.6\arcsec, and so to
compensate have larger central intensities. If there are instrumental
effects that lead to broadening, such as pointing errors, that have
not been corrected for in the data analysis of B03, then these would
cause the intensity of the central annulus to appear artificially
smaller. If there are no such effects (as suggested by the profiles of
the two nearby point sources) so that the observed intensity of the
central annulus is accurate, then our point source models would have
to be reduced in flux by a factor of about 1.8. This would lead to a
similar reduction in the halo intensities and the contribution they
make to the extended emission.

We now make a crude estimate for the contribution of the scattering
halos from sources surrounding \sga to the diffuse background inside
10\arcsec. From Figure~\ref{fig:haloint} we estimate that at 4~keV for
the $A_K=7.0$ model the scattering halo contributes about 20\% of the
PSF flux at angles 1.5-10\arcsec. B03 report that the sources inside
10\arcsec have a combined flux of $1.4\pm0.1\times 10^{-2}\:{\rm
  counts\:s^{-1}}$. Thus over the 40.9~ks of the observation, the
number of expected counts per pixel (0.492\arcsec) is about 0.1. The
$A_K=3.5$ model would predict about half of this value. These
estimates are much smaller than the 1.2 counts per pixel measured by
B03 for the strength of the diffuse background, so we conclude that
the scattering halos from the innermost GC sources make a relatively
minor contribution to the local diffuse emission.



\subsection{Time Delay of Scattered Photons\label{S:tdelay}}

An additional effect that can boost the observed intensity is the
delayed scattering of enhanced flux from an earlier flare. B03
observed that the flux at the start of their observation was rapidly
declining and three times greater than the mean value. They concluded
that a flare occurred around the time of the start of their
observation ($\gtrsim 99.34\%$ confidence).  Since flares from \sga
have been observed to occur very frequently (about once per day,
Baganoff 2003), here we estimate the effects of such flares on the
scattered halos. The time dependence of the scattered halo intensity
at various angles can help to constrain the dust and gas distribution
near the Galactic center. If a typical strong flare occurred just
before or at the start of the reported observation of quiescent
emission of B03, it could have contributed significantly to the
intensity of the diffuse emission.

\begin{figure}[h]
\begin{center}
\epsfig{
        file=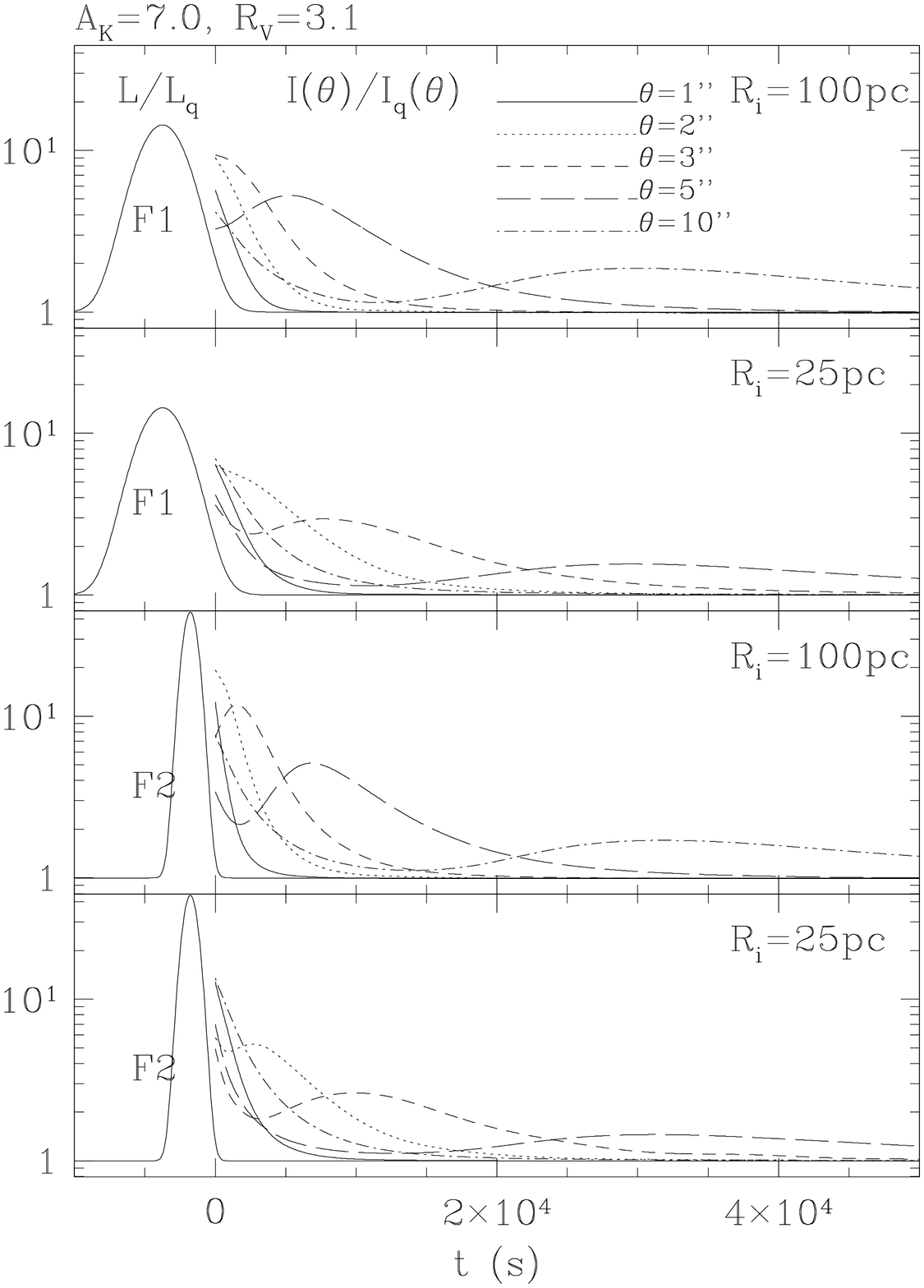,
        angle=0,
        width=4.3in}
\end{center}
\vspace*{-2.em}
\caption{
\label{fig:lightcurve2}
\footnotesize The adopted light curves of quiescent emission plus
either flares 1 (F1: top two panels) or 2 (F2: bottom two panels) are
shown by the solid lines that extend to negative times: they show the
ratio of the 2-10~keV luminosity relative to the quiescent state. The
time $t=0$ corresponds to the start of the{\it Chandra} observation of
B03. The end of the observation is approximately 40~ks later. The
flares are positioned so that the received flux is three times greater
at the start of the observation than at the end (B03). The lines that
extend from $t=0$ to later times show the evolution of the 0.5-7~keV
intensity of the scattering halo (convolved with the Chandra PSF), 
relative to the value from purely quiescent
emission, at various angles, as labeled in the figure. The 1st and 3rd
panels show the results for the standard model with the peak of the
inner component of gas and dust at $R_i=100$~pc from \sga, while the 2nd and
4th panels show the model with the same material centered only $R_i=25$~pc
from \sga. All models have $A_K=7.0$.
}
\end{figure}

\begin{figure}[h]
\begin{center}
\epsfig{
        file=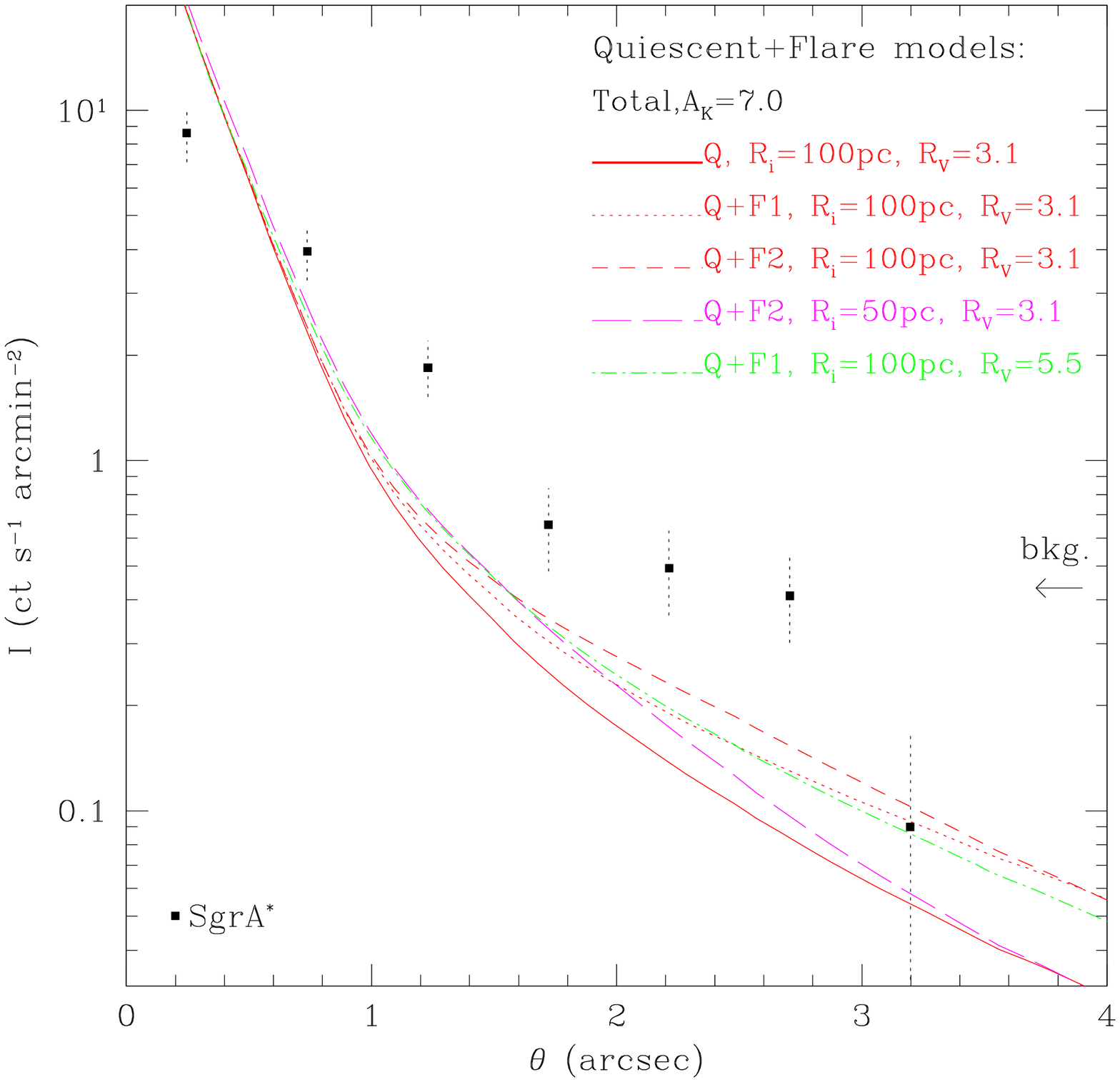,
        angle=0,
        width=4.3in}
\end{center}
\vspace*{-2.em}
\caption{
\label{fig:halonormsimp2}
\footnotesize Comparison of observed (B03) radial intensity profiles
of \sga and various theoretical profiles of quiescent
and flaring point sources, that are convolved with the Chandra PSF, and broadened 
by the effects of dust-scattering.
The parameters of the flares, F1 and F2, are listed in
Table~\ref{tab:spec}, and their hypothetical time history with respect
to the observing time of B03 is illustrated in Figure
\ref{fig:lightcurve2}.  The intensity (averaged over the $\sim 40$~ks
observation period of B03) of the quiescent+flare models is boosted
with respect to the pure quiescent models because of the delayed
arrival of X-rays from the earlier, unobserved, more luminous phase.
The mean intensity at a few arcseconds in a 40~ks observation can be
boosted by a factor of two, if a typical strong flare occurred just before the
observation.  }
\end{figure}

Photons that reach us via one or more scatterings will have
to travel a longer path length and so will be delayed compared to
those arriving directly. If the source is variable, such as a flare,
then this time delay can yield information on the distance and dust
distribution to a source (Tr\"umper \& Sch\"onfelder 1973; Draine \&
Tan 2003). For single scattering from dust that is a distance $R=yD$ along the
line of sight to a source at distance $D$, the delay is
\begin{eqnarray}
\delta t_1 (y,\theta) &=& 
\frac{D}{c}
\left\{
\frac{y}{\cos\theta} 
+(1-y)\left[1+\left(\frac{y}{1-y}\right)^2\tan^2\theta\right]^{1/2}
-1\right\}
\\
&\approx& \frac{D}{c}\frac{y}{1-y}\frac{\theta^2}{2}
= 9.7 \left(\frac{D}{8.0\kpc}\right)\frac{y}{1-y}
\left(\frac{\theta}{1\arcsec}\right)^2\:{\rm s}.
\end{eqnarray}
We expect much of the dust towards \sga to be close to the source: for $R_i=100,50,25$~pc, $1-y=0.0125, 6.25\times 10^{-3}, 3.125\times 10^{-3}$, so the delay $\delta t_1$ for $\theta=1\arcsec$ becomes
760, 1500, 3100~s.



Figure \ref{fig:lightcurve2} shows the time evolution of the scattered
halo intensity following two different hypothetical flares, whose
properties we have take to be similar to the flare observed by B01
(F1) and that observed by Porquet et al. (2003) (F2). The evolution
of intensities depends on the distribution of gas near \sga.  Given
the actual lightcurve of an observed flare, and with careful
subtraction of the quiescent component, one may use the time-varying
component of the X-ray halo to constrain the line of sight gas
distribution close to the Galactic Center.

To compare to the observations of B03 we find the mean intensity
averaged over the $\sim$40~ks observation period of B03. The results
are shown in Figure \ref{fig:halonormsimp2}. The delayed arrival of
X-ray photons from when \sga was in a more luminous state can boost
the mean halo intensity in the following $\sim$40~ks by a factor of
two or so. We stress that for these particular observations, while
there is evidence for a flare near the beginning of the observation
period, the actual lightcurve is unconstrained. Thus our calculation
makes use of representative flare properties, as seen during other
observations.

\section{Conclusions\label{S:con}}

The large column densities of gas and dust towards the Galactic center
strongly absorb and scatter X-rays, so that a careful treatment of
these processes is necessary to infer the intrinsic properties of
sources in this region, such as their luminosity, spectra, and spatial
size. The fact that much of the dust lies close to the GC means that
the scattering halos are much more concentrated than the halos seen
around typical sources in the outer Galaxy. The same phenomenon also
occurs for halos observed around X-ray sources in
star-forming regions, since these also tend to contain a large column
density of dust near the source.

In this paper we have analyzed the effects of the intervening gas and
dust on Galactic center X-ray sources, using a realistic range of
models for the total column and spatial distribution. We used
reasonable estimates for chemical abundances and depletions of the gas
and dust. We calculated the radiative transfer using realistic
scattering properties of dust grains, and considered the
effects of multiple scattering. We also evaluated the importance of
different grain size distributions.

The principal conclusions of our work are the following:

1. The intrinsic luminosities and spectra of emission from \sga in its
quiescent state and in two flaring states are shown in
Figure~\ref{fig:specqpownew} and Table~\ref{tab:spec}. These estimates depend on the amount of
intervening dust and gas, which we have related to the most relevant
observable, the extinction in the K band. The quiescent 2-10~keV
luminosity ranges from 3 to $6\times10^{33}\:{\rm ergs\:s^{-1}}$, as
$A_K$ ranges from 3.5 to 7.0 magnitudes. At the same time, the
spectral index, $\Gamma$, ranges from 1.8 to 3.3. The flaring states
are about 10 to 30 times more luminous, with the flare observed by
Chandra having a harder spectral index by about 0.7-1.0 and the flare
observed by XMM-Newton having a harder spectral index by about 2.0.
The uncertainties can be reduced once a better estimate for $A_K$ is available.

2. We show that scattering by dust within $\sim 100$~pc of the
Galactic center can account for the arcsecond-scale broadening of Galactic center
X-ray sources seen in Chandra observations by B03.

3. We have calculated the azimuthally averaged intensity profile of
the dust-scattered halo around \sga, and shown that it can
dominate the Chandra PSF beyond $\sim$1 arcsec (see Fig. 7). 
This effect contributes to the diffuse emission seen around \sga.

4. The dust-scattered halos of multiple sources in the GC region overlap, but only
account for at most $\sim 10\%$ of the diffuse emission inside
10\arcsec of \sga.


5. Motivated by the detection of enhanced emission at the start of the
observation of B03 and by other observations that show that strong
flares occur about once per day (Baganoff 2003), we have modeled the effects of flared
emission from \sga. For reasonable distributions of the dust, the time
delay for scattered emission is of order a thousand to several
thousand seconds at 1\arcsec, increasing quadratically with the
angle. We show how measurements of the evolution of the scattered halo
following a flare can help constrain the dust's spatial distribution.
It is possible that delayed scattering from a flare accounts for some of
the diffuse emission close to \sga in the observation of B03. To
disentangle the quiescent component, data in a period at least $\sim
50$~ks after a major flare needs to be analyzed.

6. Our fiducial models for the contribution of a dust-scattered halo
to the extended emission of \sga can explain up to $\sim 1/3$ of the
observed intensity at $\sim 1\arcsec$, corresponding to the extent of
the Bondi radius. Thus earlier estimates of the gas luminosity and
density may be overestimated by factors of $\sim 3/2$ and $\sim
\sqrt{3/2}$, respectively. Previous determinations of the Bondi
accretion rate and luminosity could be overestimated by similar
factors.  We conclude, like B03, that the emission around \sga is
extended.  To explain the entire emission as being due to dust
scattering would require at least a doubling of the observed column in
material at least $\sim$25~pc from \sga (but not substantially
changing the columns to the other nearby GC sources), and/or the
presence of an exceptionally strong flare just prior to the
observation period. The latter possibility can be tested by looking at
the observed halo from \sga in more quiescent periods in existing (but
unpublished) data.

\acknowledgements We thank Fred Baganoff for providing the radial
intensity profiles of several GC sources in electronic form and for
helpful explanations about the Chandra observations and data reduction.
We thank Diab Jerius, Eliot Quataert and Randall Smith for helpful discussions.
This work was supported in part by NSF grant AST-9988126,
and in part by NASA grant NAG5-10811. JCT has received
support via a Spitzer-Cotsen Fellowship from the Department of
Astrophysical Sciences and the Society of Fellows in the Liberal Arts
of Princeton University.

\end{document}